\documentclass[journal=jacsat,manuscript=article]{achemso}

\usepackage{graphicx} 
\usepackage{chemformula} 
\usepackage[T1]{fontenc} 
\usepackage{bm}
\usepackage{array}

\author{John Linderman}
\affiliation{Department of Chemistry, University of Chicago, Chicago, IL 60637, USA}
\alsoaffiliation{James Franck Institute, University of Chicago, Chicago, IL 60637, USA}

\author{Shiqi Chen}
\affiliation{Department of Chemistry, University of Chicago, Chicago, IL 60637, USA}
\alsoaffiliation{James Franck Institute, University of Chicago, Chicago, IL 60637, USA}
\altaffiliation{Current address: Department of Chemistry, University of Massachusetts Boston, Boston, MA 02125, USA}

\author{Yanzeng Li}
\affiliation{Department of Physics, Optical Engineering, and Nanoengineering, Rose-Hulman Institute of Technology, Terre Haute, IN 47803, USA
}

\author{Alexandria Hoehn}
\affiliation{Department of Chemistry, University of Chicago, Chicago, IL 60637, USA}
\alsoaffiliation{James Franck Institute, University of Chicago, Chicago, IL 60637, USA}

\author{Stuart A. Rice}
\affiliation{Department of Chemistry, University of Chicago, Chicago, IL 60637, USA}
\alsoaffiliation{James Franck Institute, University of Chicago, Chicago, IL 60637, USA}

\author{Norbert F. Scherer}
\email{nfschere@uchicago.edu}
\affiliation{Department of Chemistry, University of Chicago, Chicago, IL 60637, USA}
\alsoaffiliation{James Franck Institute, University of Chicago, Chicago, IL 60637, USA}

\title{Pseudorotation and N-body Forces in an Optical Matter System}

\begin{tocentry}
\includegraphics[scale=0.33]{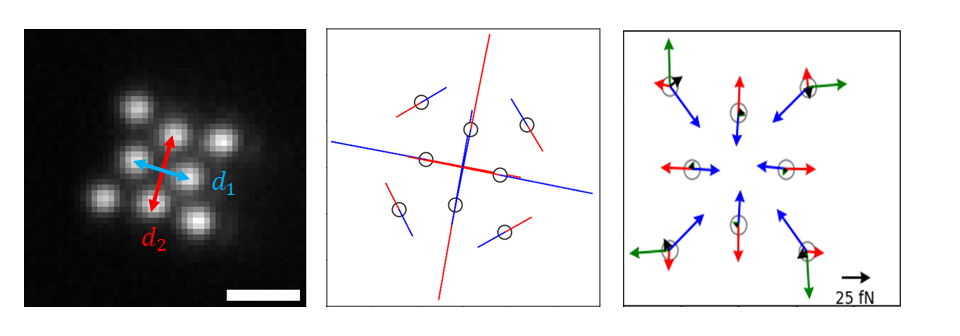}
\end{tocentry}

\begin{document}

\maketitle
\begin{abstract}
Isomerization in molecular systems almost invariably occurs through 3-dimensional motion due to the nature of chemical bonding. Pseudorotation is an unusual type of isomerization that occurs in some high symmetry systems that gives the appearance of rigid-body rotation yet only involves atom rearrangements. This paper demonstrates that pseudorotation occurs in 2-dimensions in an optical matter (OM) system of metal nanoparticle constituents. The difference in dimensionality of the dynamics arises from the electrodynamic field-interference nature of  \textit{optical binding} vs. quantum mechanical \textit{bonding} in polyatomic molecules. The 8-nanoparticle OM "kite" structure we study in experiments and simulations has $D_2$ ($D_{2h}$) symmetry and a $D_4$ symmetric transition state. The mechanism for pseudorotation involves correlated motion of all 8 nanoparticles with smooth (continuous) evolution of their interactions and without particles jumping in or out of the OM array. While the OM kite structure only occurs with 10\% probability vs. other OM isomers, its rate of pseudorotation is rapid relative to transitions to other structural isomers (e.g., "teardrop"). The other isomers have structures that lie on a trigonal lattice with interparticle separations at distances that enhance field interference and induced polarizations. Even though the kite isomer has inter-particle separations that would manifest destructive interference on a particle pair (i.e., 2-body) basis, the kite structure is the slowest to rearrange into any other isomer. We show that the unusual structure and dynamics of the kite optical matter system result from N-body interactions and forces demonstrating that N-body effects are important in this class of active matter and presumably more generally.
\end{abstract}

\newpage

\section{Introduction}

Isomerization in molecular systems almost always involves 3-dimensional motion of its constituent atoms.\cite{solomons_organic_2016} This geometric requirement results from the 3-D nature of chemical bonding in polyatomic molecules.\cite{cotton_advanced_1999,pauling_nature_2010} \textit{Pseudorotation} is a type of isomerization that involves multi-atom motions in 3-D that result in the apparent rotation of the molecule.\cite{berry_correlation_1960} It has been studied in molecules manifesting trigonal bipyramidal hybridization and $D_{3h}$ symmetry \cite{berry_correlation_1960,chen_cavity-enabled_2022,hargittai_r_2023} and square pyramidal molecules.\cite{cass_illustration_2005} Molecules with 4 atoms would constitute a minimal (complexity) threshold for 3-D pseudorotation. In this paper we experimentally demonstrate pseudorotation of an 8-nanoparticle optical matter (OM) system with motions of the nanoparticles in only 2-dimensions. Although this is not the simplest 2-dimensional OM system that can manifest pseudorotation, our comprehensive study was made possible by experimental conditions that increased the relative population of the cognate isomer; e.g., enabling extensive statistical analysis. (Note that 4-particle OM systems could undergo pseudorotation, but we have not found the diamond-shaped isomer to be sufficiently stable to allow detailed study of pseudorotation therein.)

Optical matter (OM) systems consist of dielectric or metallic particle constituents that self-organize into ordered structures due to 2-body and N-body electrodynamic interactions.\cite{mohanty_optical_2004,demergis_ultrastrong_2012,yan_guiding_2013,yan_potential_2014,nan_dissipative_2018,parker_symmetry_2025} This phenomenon is termed optical binding.\cite{burns_optical_1990,dholakia_colloquium_2010}. Their interparticle separations and configurations are governed primarily by interference of the light scattered between them creating induced-polarizations and interparticle forces. Therefore, their primary interparticle separation is given by the wavelength of the coherent beam of light in the medium in which the OM structures form.\cite{yan_potential_2014}  Furthermore, OM systems can be controlled through the characteristics (wavelength, polarization, shape, intensity, phase, etc.) of the trapping beam and the spectroscopic properties (e.g., resonance) of the nanoparticle constituents and also their collective (i.e., lattice) resonances.\cite{yan_fabrication_2015,nan_creating_2022,peterson_electrodynamic_2024}

The 8-nanoparticle OM systems studied here are constituted of 150 nm diameter Ag (or Au) nanoparticles suspended in water in a loosely focused optical trapping beam of 800 nm (vacuum wavelength). Previous studies of optical matter systems of similar numbers of nanoparticle (NP) constituents showed that these systems fluctuate between many different stable configurations (or states), analogous to molecular isomerization.\cite{yan_guiding_2013,chen_data-driven_2021} The present study focuses on the pseudurotation phenomenon manifest in the so-called "kite" isomer (state) of the 8-NP OM system, demonstrating the mechanism of pseudorotation in this 2-dimensional system. The initial and final OM isomers of the kite state manifest $D_2$ symmetry with a $D_4$ symmetric transition state. We also explore the "inter-state" transitions among different structural isomers of the 8-NP OM system and show that the relative populations of these isomers as well as the rate of intra-state pseudorotation change with ionic strength. The trends found indicate that electrostatic forces play a role in (de)stabilizing isomers and/or transition state structures in addition to the dominant electrodynamic interactions. 

The present study is complementary to our recent simulation, machine learning, and theory-focused paper\cite{chen_pseudorotation_2025} that reported on temperature-dependent and non-equilibrium properties of the 8-NP OM kite system. The present experimental studies are complemented with electrodynamics-Langevin dynamics simulations that help determine the isomerization mechanism for pseudorotation and give insights into inter-state transitions between isomers. The simulations also allow evaluating 2-body and N-body forces associated with maintaining the kite structure and contributing to pseudorotation. In addition, the high symmetry of the OM kite system manifests collective scattering of angular momentum that, in turn, causes driven rotational motion.\cite{parker_symmetry_2025,parker_optical_2020} In fact, the N-body forces cause collective rigid body rotation, which is the only net force present in the $D_4$ symmetric transition state for pseudorotation. We believe that the 2-D geometry and less restrictive spatial interactions associated with optical binding using a circularly polarized optical trapping beam vs. molecular bonding gives isomerization and pseudorotation in OM systems a geometrical flavor.\cite{damasceno_predictive_2012}

\section{Methods}
\subsection{2.1 Experimental.} Eight-particle optical matter (OM) structures were formed in and observed with an inverted microscope optical trapping setup we have described previously \cite{peterson_electrodynamic_2024}. A schematic of the experimental setup is shown in Figure S1 in the Supplementary Information. We used 150 nm diameter PVP-coated Ag nanoparticles (Nanocomposix) in a water-filled fluid cell that self-organized into OM structures in a single focused optical beam. The optical trap was formed by a circularly polarized 400 mW continuous wave beam from a Ti-Sapphire laser ($\lambda$ = 800 nm) loosely focused with a 60$\times$ water immersion objective (Nikon DC-152Q-C00-FI). OM structures built up over time as more and more particles entered the trapping field.\cite{yan_guiding_2013,han_crossover_2018} A spatially incoherent white light source was  focused through a dark-field condenser into the NP solution for dark-field microscopy imaging \cite{peterson_electrodynamic_2024}. The scattered light was collected by the 60$\times$ objective and directed to a CMOS detector (Andor, Neo) for imaging. Images were collected at a frame rate of 450 Hz. Particle positions were obtained using single-particle tracking software (Mosaic, Image J) and corrected for pixel-locking bias \cite{burov_single-pixel_2017, yifat_analysis_2017}. The closest packed OM structures were identified and sorted using a combination of adjacency matrices and trigonal lattice fitting.\cite{peterson_electrodynamic_2024}

\subsection{2.2 Simulations and Collective Coordinates.}
Electrodynamics-Langevin dynamics (EDLD) simulations \cite{sule_electrodynamics-langevin_2015} were performed using Generalized Multiparticle Mie Theory \cite{xu_electromagnetic_1995, lock_generalized_2009} (GMMT) coded in the MiePy software developed by the Scherer Lab \cite{parker_optical_2020, parker_johnaparkermiepy_2025,parker_collective_2020}. Our EDLD solver in Miepy performs a numerical integration of the Langevin equation:

\begin{equation} \label{eq1}
m\frac{d^2\bm{x}}{dt^2}=\bm{F}_{ext}(\bm{x})-\zeta\frac{d\bm{x}}{dt}+\bm{f}^{(b)}(T),
\end{equation}

\noindent
where $m$ is the mass of the particle, $\bm{x}$ is the particle's position in 3-D, $t$ is time, $\zeta=6\pi\eta r$ is the friction coefficient where $r$ is the particle radius (75 nm), $\eta$ is the viscosity of the medium (i.e., water), $\bm{f}^{(b)}(T)$ is the random (noise) force term of the thermal bath. The latter is Gaussian white noise satisfying the fluctuation dissipation relation at a specific temperature, $T$.\cite{langevin_brownian_1908,mcquarrie_statistical_2000} In the present study T = 300K. Since our OM system is overdamped on microsecond and longer timescales the left hand side is set equal to zero. $\bm{F}_{ext}$ is the external force field that consists of two parts: The first part is the electrodynamic force field computed using GMMT in MiePy.\cite{xu_electromagnetic_1995,parker_collective_2020, parker_johnaparkermiepy_2025} The second part is the electrostatic double-layer interaction between colloidal particles \cite{butt_physics_2013, chen_raman_2024}. The simulation time step is 1 $\rm{\mu}$s, which is appropriate for the overdamped dynamics of these OM systems \cite{chen_power_2024}. The dynamics of OM systems can also be represented with a set of collective coordinates.\cite{chen_data-driven_2021,chen_pseudorotation_2025} The collective motions of the NP constituents can be computed as eigenvectors by weighted principal component analysis (w-PCA) applied to particle position fluctuations obtained from electrodynamics-Langevin dynamics (EDLD) simulation trajectories \cite{chen_data-driven_2021,chen_pseudorotation_2025}.

\section{Results}

\subsection{3.1 Isomers (States) of the 8-particle OM system.}
Interparticle distances and interactions are governed by both electrodynamics, through the scattering of the incident beam by the nanoparticle constituents of the OM system, and electrostatics arising from the PVP coating on the nanoparticles that creates a charged surface and electrostatic repulsion between particles.\cite{chen_cavity-enabled_2022} The electrodynamics of an OM system can be controlled by changing constituents (nanoparticle material, size, shape) or by changing the beam conditions (polarization, intensity profile, phase profile). Adjusting the ionic strength of the colloidal nanoparticle solution changes the Debye screening length\cite{debye_electrolytes_1923} and electrostatic interactions the nanoparticle constituents experience. Increasing the ionic strength of the colloidal nanoparticle solution by adding ions (in our case NaCl) decreases the Debye screening length and reduces the electrostatic repulsion between NPs. Another consequence of increased ionic strength is the distance between the trapping plane (The 2-D plane containing the OM system) and the glass coverslip surface that is the top of the sample cell will decrease. \cite{parker_collective_2020} The electrostatic and hydrodynamic\cite{kurzthaler_particle_2020} interactions with the glass coverslip will affect the dynamics of the OM system\cite{yifat_facile_2021}; we expect lateral friction to increase and motions to decrease with increasing ionic strength. Minimizing the surface roughness of the glass coverslip and/or the inhomogenous distribution of electric charge\cite{gupta_nanoscale_2000,behrens_charge_2001,liau_ultrafast_2001} could mitigate any spatially inhomogeneous friction effect on the motion and dynamics of OM systems. 

Figure 1a-d shows representative dark-field microscopy images of different 8-nanoparticle OM isomers that form spontaneously at various ionic strengths of the aqueous solutions under the relatively loose focusing conditions vs. what is typically used single particle trapping experiments.\cite{neuman_optical_2004} Figure 1e shows the different isomers have different relative stabilities as established by their probabilities of occurrence in the many experimental images ($\sim 70,000$ frames/images) from multiple dark-field microscopy videos obtained for each specific ionic strength condition. Beam power (i.e., power density) and beam size also affect the relative stabilities. For example, changing the trapping beam from a tightly focused condition to a beam with larger width increased the probability of the OM array being in the kite isomer by nearly 2 orders of magnitude. Therefore, the results presented in this paper are for simple focused Gaussian beam conditions that approximately optimize formation of the 8-NP kite OM structure with significant probability ($\sim10\%$). 

Figure 1e shows that the "teardrop" structure \cite{han_crossover_2018,parker_optical_2020} is the most probable. In fact, this is true for all beam conditions and ionic strengths explored, particularly in tightly focused beams (results not shown). See the Supplementary Information for additional isomer probability distributions for different ionic strengths (from 18 M$\Omega$ water to 1.2 mM NaCl). Note that the names of the structures follow those published for colloidal fluids and also found in acoustic trapping studies \cite{meng_free-energy_2010,lim_cluster_2019}. We are coining the "Sphinx" term here.

\begin{figure}[H]	
	\centering
	\includegraphics[width=1\textwidth]{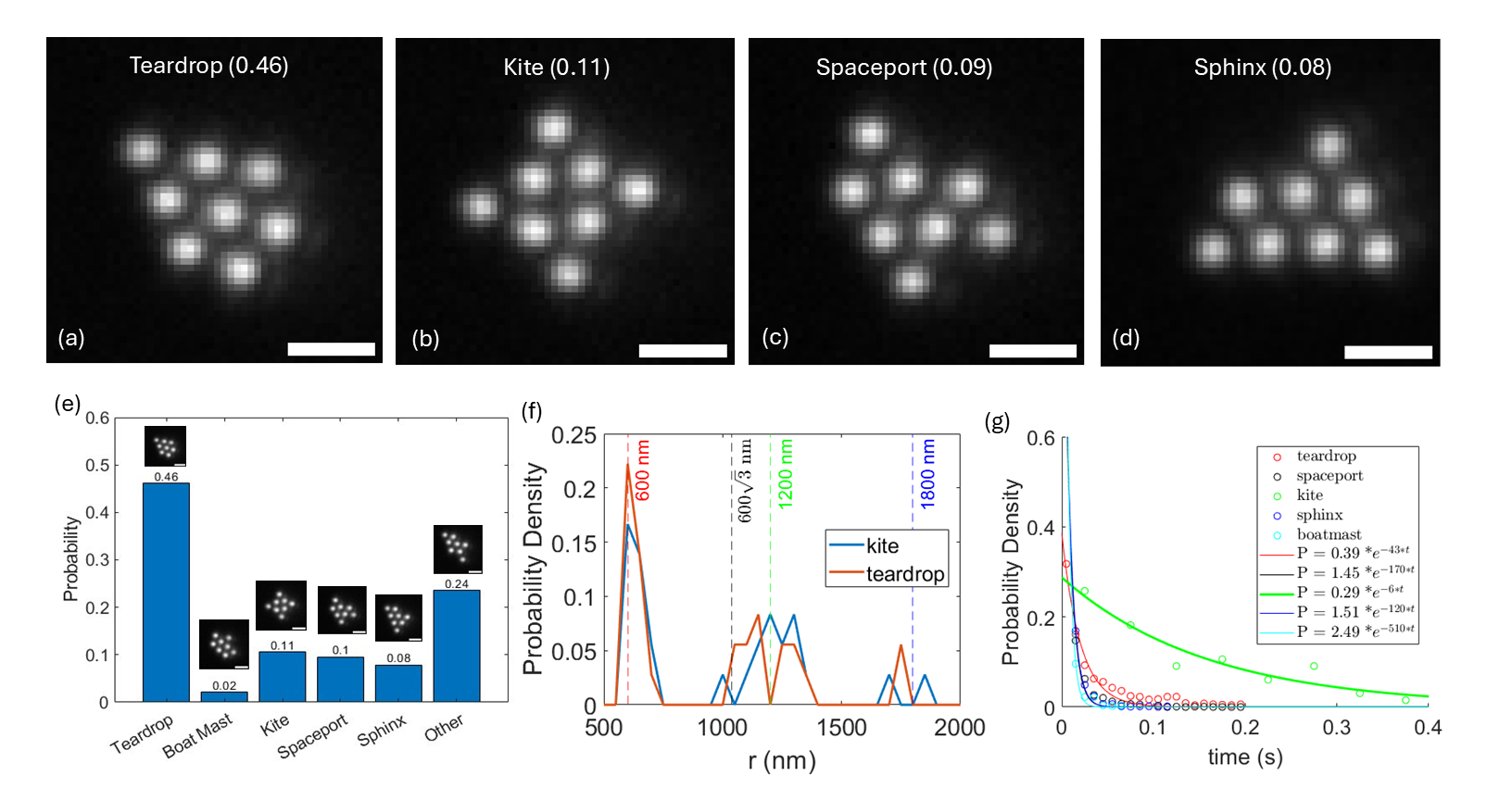}
	\caption{Darkfield microscopy images, pair distribution functions, and probabilities of different 8-NP OM isomers.  (a-d) Dark field microscopy images of the 150 nm diameter spherical Ag nanoparticle-based OM array formed in a 0.6 mM NaCl solution. (e) Probability of the 8 particle OM structure being in the “teardrop”, “boat mast”, “kite”, “spaceport”, "sphinx", or any “other” configuration. (f) Probability density distributions of the pairwise distances of particles in the "teardrop" and "kite" isomers. The distributions were collected from trajectories of particle positions obtained from dark field microscopy videos of the 8-particle OM system ($\sim$70,000 total frames). The kite data set consists of 7,497 images and 66 transitions and the teardrop data set consists of 32,736 frames and 1230 transitions. Pairwise distances of the particles can exceed 2000 nm, but the plot axes have been limited to highlight regions of interest. The bin width for the pair distribution is 50 nm. (g) Distributions of first passage times for the isomers of the 8-particle OM system to transition to another isomer. The bin width for the first passage time distributions is 50 ms for "kite" and 10 ms for all others. The difference in bin sizes is because there are fewer kite trajectories (66) than trajectories for the other isomers (>500 each). Reducing the bin width 2-fold for the kite isomer 1st passage time data had no effect on the fitted rate constant.}
	\label{fig:1}
\end{figure}

Figure 1e shows the pair distribution functions for the kite and teardrop OM states. See the Supplementary Information for all 4 distributions. All 4 distributions have a peak at 600 nm, which is the characteristic optical binding distance for the OM system. The center particles of the kite isomer do not have this separation, but the 3 particles in each corner do.  The next peak of interest is at $600\sqrt{3} \approx 1050$ nm. The three isomers with particles on a trigonal lattice all manifest this interparticle separation, but the probability density of the kite structure dips to zero at this value. All four structures have peaks at 1200 nm, which is 2 optical binding distances. The kite isomer has contrarian probability density in the 900-1000 nm and 1150 nm regions to that of the trigonal lattice isomers. This probability density is associated with the separations of the inner 4 nanoparticles. All structures also have a peak at 1800 nm, which is three optical binding distances. In the case of the kite isomer this corresponds to diagonal distances between particles at the corners. See the Supplementary Information for additional details.

Figure 1f shows exponential fits to first passage time distributions associated with each isomer transitioning to any other isomer. The rate of leaving the kite isomer (green) is the slowest indicating the largest activation energy. We will return to this later in the paper. The teardrop isomer is second slowest. This is likely due to its greater stability since it is unique among 8-nanoparticle OM isomers in 2-D having 14 interparticle separations at 600 nm (800 nm / 1.33) (i.e., the first optical binding distance) which is at least one more than any other isomer. Since these first optical binding distances are the smallest separation in the optical matter structures and contribute strongly to constructive interference of the EM fields at the location of the nanoparticles, they result in greater induced polarization and stronger optical binding \cite{peterson_electrodynamic_2024}.

\subsection{3.2 Intra-state transitions of the OM kite isomer: the $d_1$ - $d_2$ indicator of  Pseudorotation.}
Pseudorotation motion of the kite isomer of the 8-particle OM system was initially observed in experiments and subsequently found in EDLD simulations. The present comprehensive experimental study of the pseudorotation phenomenon was made possible by the aforementioned optical trapping conditions that dramatically increase the probability of the 8 particle OM system being in the kite isomer. 

Figure 2a-c show three consecutive frames (images) of a dark-field microscopy video of the 8 particle OM system demonstrating pseudorotation (see Supplementary Video 1). The double-headed colored arrows superimposed in Figure 2a define the distances $d_1$ and $d_2$ associated with the orthogonal pairs of the inner 4 NPs. We use the difference of these two distances, $d_1$ - $d_2$, as an indicator of pseudorotation. 

Figure 2d shows the time-dependent difference of the $d_1$ and $d_2$ distances of the orthogonal pairs of the four inner particles of the OM kite isomer, which are the primary component of the pseudorotation motion of the system as a function of time. The trajectory of $d_1$ - $d_2$  in Figure 2d is the OM system in the kite isomer during a portion of a longer (nearly 1000 frames) dark-field microscopy video. The duration or first passage time of being in the $d_1 < d_2$ (or $d_1 > d_2$) the kite isomer varies stochastically; a majority of kite isomers persist for less than 1 second (the first passage time distributions are shown in Figure 3f). Figure 2e shows a trajectory of the $d_1$ - $d_2$ separation of the 8-particle OM system from an EDLD simulation for 100 mW beam power and 50 nm Debye screening length. The trajectory in Figure 2e consists of 2 million 1 $\mu$s duration timesteps binned and averaged to match the 2.2 ms experimental timestep (450 fps). The trajectories from experiment (Figure 2d) and simulation (Figure 2e) are qualitatively similar, but the range of $d_1 - d_2$ is smaller in the simulation results due to averaging. See Supplementary Information for additional details.

Figure 2f shows probability density representations of particle positions for the experimental trajectory shown in Figure 2d, while Figure 2g shows the probability density of particle positions for the trajectory from Figure 2e. The results from simulation and experiment are very similar. Note that since the collectively scattered light from the symmetrical kite isomer carries (off) orbital angular momentum, the isomer undergoes rigid body rotation. This motion is subtracted when creating Figures 2f,g. See the Supplementary Information (Figure S4) for additional results related to Figures 2d and 2g with various beam powers and Debye screening lengths that we used to establish EDLD simulation conditions that closely match the experimental conditions.

\begin{figure}[H]	
	\centering
	\includegraphics[width=1\textwidth]{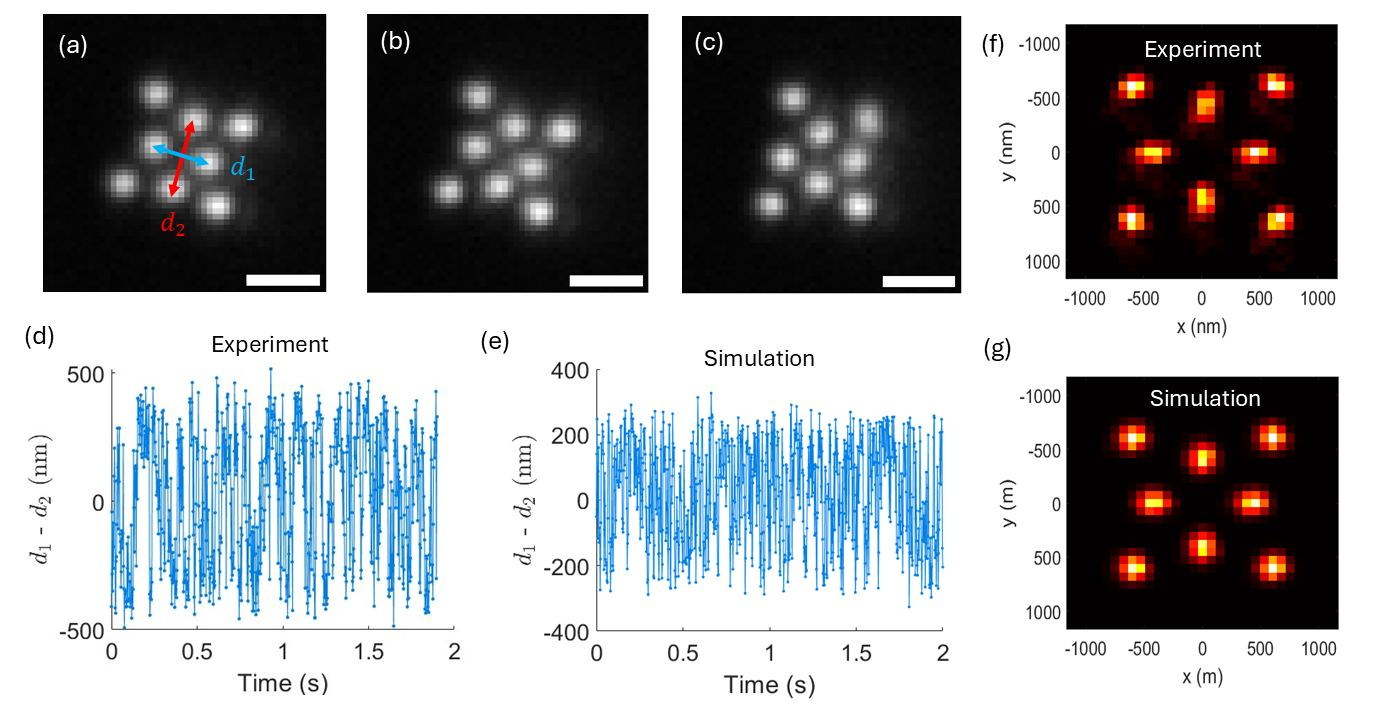}
	\caption{Experimental measurements and EDLD simulations characterizing the dynamics of the 8 Ag nanoparticle OM kite isomer. (a) - (c) Dark field microscopy images of an 8-particle OM array of 150 nm diameter spherical Ag nanoparticles. The blue and red arrows in (a) represent the two distances (labeled $d_1$ and $d_2$) between orthogonal pairs of central particles of the OM array. The three panels from consecutive frames (450 fps) display pseudorotation. The scale bars are 1 $\mu$m. (d) A single experimental time trajectory of the difference of the two distances ($d_1 - d_2$). The OM array is in the kite isomer for the duration of the trajectory and is in a 0.6 mM NaCl solution. (e) A single EDLD simulated time trajectory of $d_1 - d_2$ for an 8 Ag nanoparticle array for 100 mW beam power and 50 nm Debye screening length. The array persists in the kite isomer (state) for the duration of the trajectory. (f) Probability density of rotation-reset particle positions for the time trajectory from experiment shown in  (d). (g) Probability density of rotation-reset particle positions for the time trajectory from the simulation shown in (e). Bins are 75x75 nm in size for both probability density plots. The bin size essentially matches the effective pixel size of the experimental dark field microscopy images.}
	\label{fig:2}
\end{figure}

\subsection{3.3 The effect of ionic strength on the intra- and inter-state dynamics of
the kite structure.}
Figure 3a-c show scatter plots of the $d_1$ and $d_2$ values obtained from experiment for different ionic strengths of the solution; from 18 M$\Omega$ pure water to 0.6 and 1.2 mM NaCl solutions. The scatter plots manifest a bimodal character. Best linear fits are shown in red, with slopes of nearly -1 demonstrating a strong anti-correlation of the separations of the orthogonal pairs of central particles. These results agree with analogous distributions for simulations performed at 300K and 27 nm Debye screening lengths.\cite{chen_pseudorotation_2025} 

Figure 3d shows the projections of the distributions of Figure 3a-c onto their best fit lines. The results are clearly bimodal, but some also manifest an asymmetry of the probabilities of the $d_1$ < $d_2$ and $d_1$ > $d_2$ configurations. The asymmetry of the bimodal distributions seen in Figure 3d results from: (i) our convention to make the $d_1$ and $d_2$ assignments $d_1$ < $d_2$ for the first configuration in each trajectory that our data analysis code determines as a kite structure; (ii) the comparable time scales of the distributions of first passage times of the intra-state transition and the duration of a given trajectory of the kite structure; (iii) and that other motions (fluctuations) particularly of the outer 4 nanoparticles are also involved in pseudorotation. The first two points are considered in this section and the Supplementary Information, while point (iii) is considered in the next section and also the Supplementary Information.  

When the trajectory lengths of each epoch of the OM kite state approach the time scales of the pseudorotation transitions, the bias present in the definition of $d_1$ and $d_2$ becomes relevant. In contrast, the $d_1$ - $d_2$ distribution associated with the trajectory shown in Figure 2d, which contributes to the green curve in Figure 3d, indicates that each configuration is equally likely. The time trajectory shown in Figure 2d is long with more than 800 timesteps and scores of intra-state transitions. Therefore, the choice of the initial condition as $d_1$ < $d_2$ would only cause a small, almost negligible bias. In fact, when we  randomly assign the initial configuration to be either $d_1 < d_2$ or $d_1 > d_2$ the probability density distributions become symmetric.

We complement this probability distribution analysis with analysis of the dynamics between the kite isomers using experimental $d_1 - d_2$ trajectories like that of Figure 2d to calculate first passage times for the forward and backward \textit{intra-state} transitions of the OM kite isomer. Since the initial and final states are identical modulo pseudorotation, forward is defined where the initial configuration of a given portion of a trajectory has configurations with $d_1$ < $d_2$ prior to a transition. Exponential distributions are found and fitted for both forward ($d_1$ < $d_2$ $\rightarrow$ $d_1$ > $d_2$) and reverse ($d_1$ > $d_2$ $\rightarrow$ $d_1$ < $d_2$) processes. Figure 3e shows the first passage time distribution of forward intra-state transitions (pseudorotation) of the kite system obtained from a time trajectory of $d_1$ - $d_2$ interparticle distance fluctuations. The simulated fluctuations of the 8 particle OM system along the pseudorotation coordinate are similar to experimental results (Figure 2d). See the Supplementary Information for additional details.

The ionic strength of the medium has a noticeable effect on the first passage time distributions of intra-state $d_1$ - $d_2$ transitions (Figure 3e). The results from simulation indicate a positive correlation between first passage time and Debye screening length; i.e., the first passage time distributions decay more quickly for shorter Debye screening lengths. Figure 3f shows that the lifetime of a  OM structure in the kite isomer is also affected by ionic strength. The OM kite isomer is more probable at moderate ionic strength (and the associated reduced Debye screening length) vs. 18 M$\Omega$ water as suggested by the greater number of data points in the distribution, hence realizations of the kite isomer. Therefore, the kite isomer is stabilized in the 0.6 $\mu$M NaCl condition. Finally, a comparison of Figures 3e,f show that the \textit{intra-state} pseudorotation transitions of the OM kite isomer occur on times scales faster (3X to over an order of magnitude) than the \textit{inter-state} transitions from the kite isomer to other isomers (Figure 3f).

\begin{figure}[H]	
	\centering
	\includegraphics[width=1\textwidth]{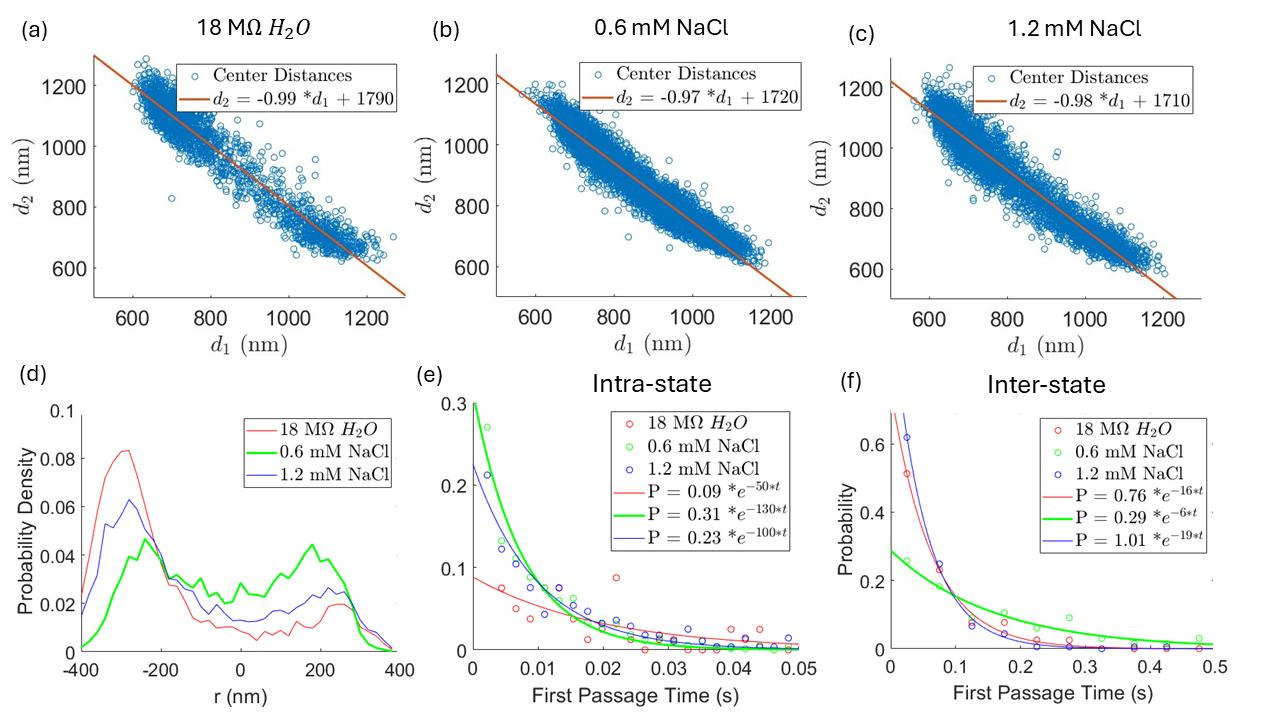}
	\caption{Ionic strength affects the stability and kinetics of the OM kite isomer. (a) - (c) Scatter plots from experiments (videos) of the center distances of 8-particle OM arrays in the kite isomer obtained from time trajectories taken in varying ionic strength solutions (18 M$\Omega$ H$_{2}$O (2,181 frames out of 50,000 total), 0.6 mM NaCl (7,496 frames out of 71,007 total), 1.2 mM NaCl (4,644 frames out of 65,446 frames). The red lines represent a linear regression of the scatter plots. The regressed lines are given by $r=(d_1+md_2)/\sqrt{m^2+1}$ where $m=-0.99$ is the slope. (d) Probability densities of the projection of the scatter plots onto their corresponding fit line. (e) Distribution of first passage times for the kite structure to switch from $d_2$ <$d_1$ (defined as the initial condition for any new kite structure) to $d_2$ > $d_1$. (f) Distribution of the first passage times of an individual OM kite structure transitioning to any other isomer for the three different ionic strength solutions. The shorter lifetimes for the 18 M$\Omega$ H$_{2}$O and the 1.2 mM NaCl cases contribute to causing the bias in the associated probability densities in (a), (c), and (d) toward the defined initial condition ($d_1$<$d_2$).  }
	\label{fig:3}
\end{figure}

\subsection{3.4 Reaction coordinate for pseudorotation of the 8-particle OM kite isomer.}

It is important to have a quantitative characterization of the reaction coordinate for the intra-state isomerization of the 8-NP OM kite system to objectively establish that it is a pseudorotation. By symmetry the transition state of pseudorotation would be the rotationally symmetrical OM kite structure with a $D_4$ point group where $d_1 = d_2$ and therefore in the $r=0$ region of Figure 3d. An example of this symmetrical structure is shown in Figure 2b. We quantify this intuitive suggestion that this symmetrical structure is the transition state for pseudorotation in this section.

We have previously used a committor analysis to validate the choice of a reaction coordinate for isomerization of a different optical matter system.\cite{chen_data-driven_2021} Here we use a committor analysis to determine the importance of $d_1$ - $d_2$ as (part of) a reaction coordinate. The committor analysis performed involved sampling 686 8-particle configurations of the kite isomer taken from an experimental trajectory. These configurations were the initial conditions for ED-LD simulations. We generated 200 simulation trajectories from each initial condition and evaluated the committor function by counting the trajectories that passed a defined threshold ($d_1$ - $d_2$ = $\pm$ 250 nm) to either the reactant ($d_1$ < $d_2$) or product ($d_1$ > $d_2$) states within the calculated trajectory length (2000 time steps of 1 $\mu$s step duration). The committor analysis shows that kite structures with $d_1 \simeq d_2$ values are almost equally likely to go toward either state defined by $d_1 < d_2$ and $d_1 > d_2$. In contrast, if the initial configurations are already asymmetric structures and closer to one of the peaks in the bimodal probability density distribution of Figure 3d, then that initial configuration is less likely to make a transition and instead remain near that peak. The results shown in Figure 4a support the importance of $d_1$ - $d_2$ as a (portion of a) reaction coordinate for the pseudorotation of the 8-particle OM structure.

\begin{figure}[H]	
	\centering
	\includegraphics[width=.7\textwidth]{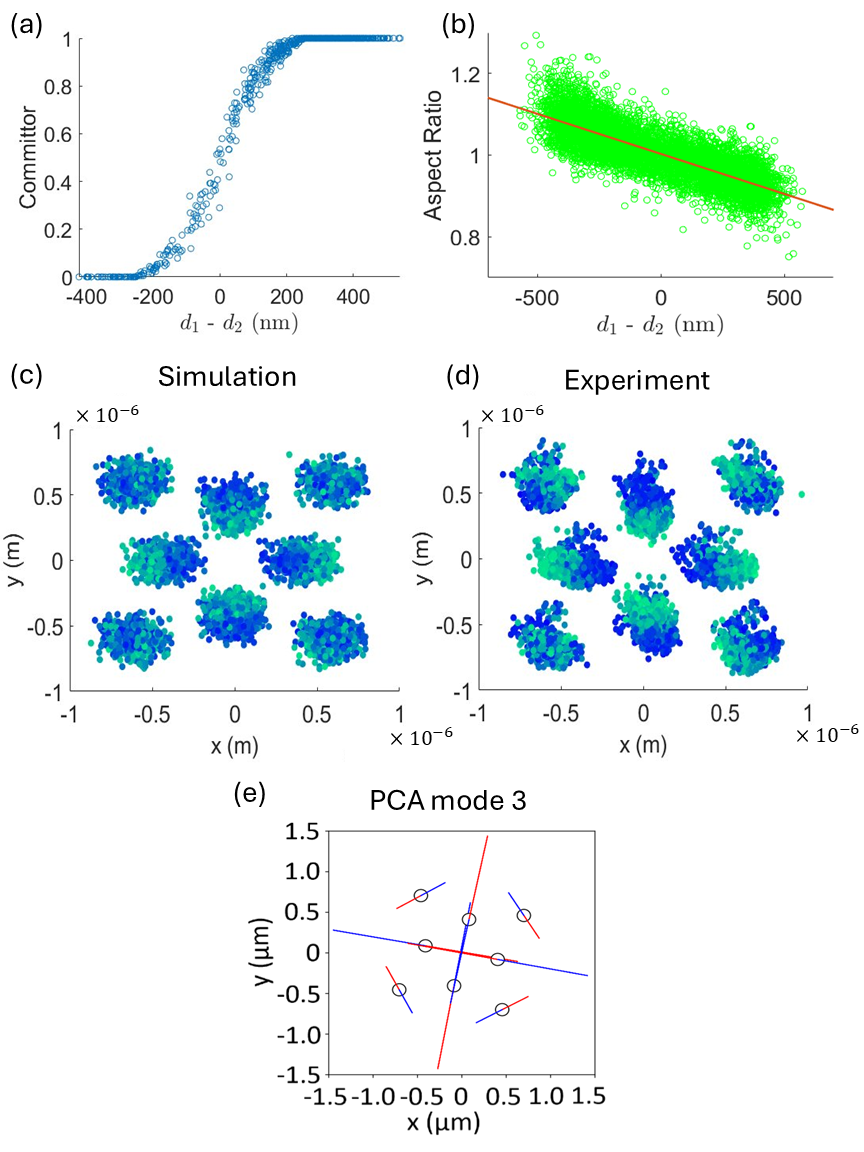}
	\caption{The reaction coordinate for pseudorotation includes the $d_1 - d_2$ parameter. (a) Plot of committor probability for 686 configurations selected from experiment along the $d_1$ - $d_2$ coordinate. The configurations were selected from experimental video data of a single continuous kite trajectory (duration of 686 frames). (b) Scatter plot of the aspect ratio of the sides of the quadrilateral formed by the outer four particles of the kite structure vs. $d_1 - d_2$. The orange line represents a linear regression of the scatter plot, whose function is $AR = m(d_1 - d_2) + 1$, where $m = 1.96 \times 10^{-4}$. (c) Plot of rotation-reset particle positions of randomly selected timesteps of a single EDLD simulated time trajectory. (d) Rotation-reset particle positions of a single time trajectory of experimental dark field microscopy images. These color-coded scatter plots condition particle positions on the value of $d_1 - d_2$. The green (blue) points are the rotation-reset particle positions of kite isomers where $d_1 > d_2$ ($d_1 < d_2$). (e) Mode 3 of the PCA modes for the kite structure.\cite{chen_data-driven_2021,chen_pseudorotation_2025} The red/blue colors indicate the phase of the collective motion and the length of the line segment is the standard deviation of the motion multiplied by 200. This mode is the collective motion of pseudorotation. }
	\label{fig:4}
\end{figure}

While the committor analysis demonstrates that $d_1$ - $d_2$ is an important part of a reaction coordinate for intra-state pseudorotation isomerization of the 8-NP OM kite system, it is reasonable to assume that there must be some motion of the outer 4 NPs that is likely correlated to the $d_1$ - $d_2$ motion.  Figures 2f,g make clear that all 8 NPs move, but this result does not demonstrate if or how their motions are correlated. Insight is provided by the aspect ratio of the quadralateral defined  by the outer four particles. Figure 4b shows that there is an (anti-)correlation between the shape (i.e., aspect ratio) of the quadrilateral formed by the outer 4 NPs and the $d_1$ - $d_2$ values of the inner 4 NPs. (Note that the slope of the plot in Fig. 4b would be positive if we defined the aspect ratio in the $\pi/2$ rotated orientation)  

Figure 4c,d shows scatter plot representations of the positions of all 8 NPs in the kite state conditioned on their $d_1 - d_2$ value obtained from experimental (Fig. 4d) and EDLD simulation (Fig. 4c) results shown in Figures 2f and 2g, respectively. Specifically, the color gradient representation shows the positions of the 8 nanoparticles conditioned on the value of $d_1 - d_2$; therefore points of the same or similar colors correspond to similar kite structures. The green (blue) points are the rotation-reset particle positions of kite isomers where $d_1 > d_2$ ($d_1 < d_2$). Figures 4c,d clearly show correlated motions of the outer 4 particles with the inner 4 particles, and support the conjecture that correlated motions of all 8 nanoparticles constitute pseudorotation. See the Supplementary Information for additional results and details.

This correlated motion is captured in Figure 4e, which shows a collective mode of the 8-NP OM kite system obtained from the principal component analysis (PCA) of the fluctuations of the nanoparticles in the kite isomer. \cite{chen_data-driven_2021,chen_pseudorotation_2025} The red/blue colors indicate the phase of the collective motion and the length of each line segment is the standard deviation of the motion of the associated nanoparticle multiplied by 200 for visual clarity. The collective motion is termed PC mode 3 that captures the $d_1 - d_2$ motion of the inner 4 nanoparticles and the correlated motion of the outer 4 nanoparticles. \cite{chen_pseudorotation_2025}. Note that the (anti-)correlation of the aspect ratio with  $d_1$ - $d_2$ is consistent with the motion described by PCA mode 3, giving experimental support to the PCA obtained from simulation. Together, the results shown in Figure 4 show that the collective motion represented by PCA mode 3 represents the reaction coordinate for pseudorotation.

\subsection{3.5 Forces associated with kite isomer stabilization and at the transition state for pseudorotation}

Conservative, non-conservative and non-reciprocal forces arise in optical matter (OM) systems. For example, we recently showed that  nonconservative N-body forces arise due to spatial symmetry breaking in 3 or more particle systems even for identical nanoparticle constituents\cite{parker_symmetry_2025}. Therefore, it is important to consider the multiparticle scattering contribution to the local field and field gradients in determining the force acting on individual particle constituents in OM systems. In this section we develop an understanding of the forces arising in the kite optical matter system and the effect of $D_2$ or $D_4$ symmetry of the kite OM isomer vs. the cylindrical symmetry of the circularly polarized incident electromagnetic field (i.e., trapping beam). Our analysis compares the total force and decomposition thereof acting on the nanoparticle constituents to different 1-, 2- and N-interaction (c.f., N-body) contributions. Section 10 in the Supplementary Information gives a more detailed derivation of the equations presented in this section.

In the point dipole approximation, the time-averaged force of a particle with an induced polarization (i.e., induced dipole), $\boldsymbol{p}$, in the $\boldsymbol{q}$ direction is\cite{dholakia_colloquium_2010,chaumet_physical_2000,novotny_principle_2012,parker_symmetry_2025}
\begin{equation}
    \left\langle {F}_{{\boldsymbol{q}}}\right\rangle=\frac{1}{2}\text{Re}\left[{{\boldsymbol{p}}}^{ * }{\partial }_{{\boldsymbol{q}}}\left({{{\boldsymbol{E}}}}\right)\right] \label{start}
\end{equation}
where $\langle...\rangle$ denotes a time (i.e., cycle) average, $\boldsymbol{q}$ represents  the x,y,z directions, $\boldsymbol{E}$ is the total electric field at the location of the particle, ${\partial }_{\boldsymbol{q}}$ denotes a partial derivative in the $\boldsymbol{q}$ direction, and an asterisk denotes the conjugate transpose. Assuming spherical nanoparticles with an isotropic (scalar, but complex) polarizability, $\alpha$, the induced-polarization is directly proportional to the electric field at the location of the particle; i.e., $\boldsymbol{p} = \alpha\boldsymbol{E}$ \cite{novotny_principle_2012}. Under these conditions, the induced-polarization of a particle labeled by subscript j in the presence of a collection of other particles located in a 2-dimensional transverse plane can be written as\cite{parker_symmetry_2025,lakhtakia_macroscopic_1990}
\begin{equation}
    \begin{aligned}
          \boldsymbol p_j=\alpha\
            \mathbf E_0(\mathbf r_j) +\alpha^2 \sum_{k\neq j}\overline{\mathbf G}_{r_j r_k}\,\mathbf E_0(\mathbf r_k)
            + \alpha^3 \sum_{k\neq j}\sum_{l\neq k}\overline{\mathbf G}_{r_j r_k}\,\overline{\mathbf G}_{r_k r_l}\,\mathbf E_0(\mathbf r_l)+  \mathcal{O}(\alpha^4)+ \dots\label{p}  
    \end{aligned}
\end{equation}
where $\mathbf E_0$ is the incident field and $\overline{\boldsymbol{G}}_{{r}_{j}{r}_{k}}$ is the dyadic Green’s function  \cite{dholakia_colloquium_2010} that propagates an electric field radiated by an induced polarization at position $\boldsymbol{r}_k$ to the position $\boldsymbol{r}_j$. The particle identities are denoted using j,k,l, etc.  Similarly, the field gradient in the $\boldsymbol{q}$ direction at the position $\boldsymbol{r}_j$ for an incident plane wave can be written as\cite{parker_symmetry_2025}
\begin{equation}
    \begin{aligned}
            \partial_{\boldsymbol{q}}\mathbf {E({r_j})} = \alpha\sum_{k\neq j}\partial_{\boldsymbol{q}}\!\big(\overline{\mathbf G}_{r_j r_k}\big)\,\mathbf E_0(\mathbf r_k) + \alpha^2 \sum_{k\neq j}\sum_{l\neq k} \partial_{\boldsymbol{q}}\!\big(\overline{\mathbf G}_{r_j r_k}\big)\, \overline{\mathbf G}_{r_k r_l}\,\mathbf E_0(\mathbf r_l) \\ + \alpha^3 \sum_{k\neq j}\sum_{l\neq k}\sum_{m\neq l} \partial_{\boldsymbol{q}}\!\big(\overline{\mathbf G}_{r_j r_k}\big)\, \overline{\mathbf G}_{r_k r_j}\,\overline{\mathbf G}_{r_l r_m}\,\mathbf E_0(\mathbf r_m) + \mathcal{O}(\alpha^4)+\dots \label{de}
    \end{aligned}
\end{equation}
Substituting Eqs. \ref{p} and \ref{de} into Eq. \ref{start} leads to the following expression for the time-averaged force on particle j in direction $\boldsymbol{q}$; 
\begin{multline}
        \left\langle \boldsymbol{F}_{\boldsymbol{q}}\right\rangle_j
        =\tfrac{1}{2}\text{Re}\Big[\alpha^{*}\,\mathbf E_0(\mathbf r_j)^{*}\!\cdot\partial_{\boldsymbol{q}}\mathbf E_0(\mathbf r_j) \\ 
        +(\alpha^{*})\,\alpha\;\sum_{k\neq j}\mathbf E_0(\mathbf r_j)^{*}\!\cdot\!\partial_{\boldsymbol{q}}\overline{\mathbf G}_{r_j r_k}\,\mathbf E_0(\mathbf r_k)+(\alpha^{*})^{2}\,\sum_{k\neq j}\partial_{\boldsymbol{q}}\mathbf E_0(\mathbf r_j)\!\cdot\!\overline{\mathbf G}_{r_j r_k}^{\,*}\mathbf E_0(\mathbf r_k)\, \\
        +(\alpha^{*})^{2}\,\alpha\;\Bigg[
        \sum_{k\neq j} \overline{\mathbf G}_{r_j r_k}^{\,*} \mathbf E_0(\mathbf r_k)^{*}\times\sum_{l\neq j}\,
        \partial_{\boldsymbol{q}}\overline{\mathbf G}_{r_j r_l}\,\mathbf E_0(\mathbf r_l)\Bigg] +\mathcal{O}(\alpha^4)+\cdots)\Big] .\label{total}
\end{multline}
Eq. \Ref{total} is factorized in terms of powers of $\alpha$ allowing explicit characterization in terms of the rank of the matter-radiation interaction; i.e., 1-interaction for $\alpha$, 2-interactions for $\alpha^2$, etc.. This expression clarifies our usage of 1-interaction, 2-interaction, and N-interaction terminology as opposed to a more conventional 1-body, 2-body nomenclature, and naturally encompasses ABA-type interactions of 2 nanoparticles, but at 3rd-order in $\alpha$.

Using the identity $|\boldsymbol{E}_0|^2=\boldsymbol{E}_0^*\cdot\boldsymbol{E}_0$, differentiating and other steps given in the Supplementary Information allow rewriting  the first term with respect to $\alpha$ in the commonly expressed form,\cite{chaumet_physical_2000,novotny_principle_2012}
\begin{equation}
    \boldsymbol{F}_\text{{self}}(\boldsymbol{r}_j)
= \frac{1}{4}\,\text{Re}(\alpha)\,\partial_{\boldsymbol{q}}\big|\mathbf E_0(\mathbf r_j)\big|^2
\;-\;\frac{1}{2}\,\text{Im}(\alpha)|\mathbf E_0(\mathbf r_j)^2|\partial_{\boldsymbol{q}}\phi
\end{equation}
where the first term is the conservative intensity gradient force and the second is the non-conservative scattering force\cite{novotny_principle_2012}. The latter is also known as the phase gradient force, which is associated with the radiation pressure\cite{novotny_principle_2012}. 

We compute the total force, $\boldsymbol{F}_{\text{total,j}}$, on each (nano)-particle constituent, j, in an OM system (using the MiePy code developed previously in the Scherer group\cite{parker_johnaparkermiepy_2025}) by evaluating the Maxwell Stress Tensor with all particles at particular positions in the incident (e.g., Gaussian or plane wave) beam. Since the force is computed by matrix inversion, the term-by-term contributions to the total force are not revealed. In the following, we compute the first and second order (i.e., 1- and 2- interaction) contributions to the force of a single particle to the total to understand the forces on each particle arising from their interactions with their neighbors. Comparing these contributions to the total force allows inferring the higher order contributions.

In a Gaussian beam, the first order (i.e., 1-interaction) term is the intensity gradient force on the particle that pulls it towards the Gaussian beam's center when the nanoparticles are at the focal plane; i.e., where there is no transverse gradient of the phase since the curvature of the wavefront is 0. The forces associated with the second order term (i.e., 2-interaction) of Eq. \Ref{total} arise from pairwise interactions of nanoparticles in the beam in addition to each particle's interaction with the beam. Therefore, we define $\boldsymbol{f}_{jk}$ as the force on particle j as a 2-particle array with particle k; (i.e., a dimer).
\begin{equation}
    \begin{aligned}
        \boldsymbol{f}_{jk}
        =\tfrac{1}{2}\text{Re}\Bigg[\alpha^{*}\,\mathbf E_0(\mathbf r_j)^{*}\!\cdot\partial_{\boldsymbol{q}}\mathbf E_0(\mathbf r_j)+(\alpha^{*})\,\alpha\;\mathbf E_0(\mathbf r_j)^{*}\!\cdot\!\partial_{\boldsymbol{q}}\overline{\mathbf G}_{r_j r_k}\,\mathbf E_0(\mathbf r_k)\\+(\alpha^{*})^{2}\,\mathbf E_0(\mathbf r_k)^{*}\!\cdot\!\overline{\mathbf G}_{r_k r_j}^{\,*}\,
        \partial_{\boldsymbol{q}}\mathbf E_0(\mathbf r_j)\Bigg] .
    \end{aligned} \label{alpha2}
\end{equation}
$\boldsymbol{f}_{jk}$ is computed using the first two terms in the expansions of the matter-radiation interaction given in Eq. \Ref{total}. The intensity gradient term is nonzero in a Gaussian beam, so $\boldsymbol{f}_{jk}$ includes both the single particle (1-interaction) force and the force between two particles (2-interaction). Therefore, the 2-interaction force, $\boldsymbol{F}_j^{(2)}$, on particle j is the sum of the forces of all possible 2-particle arrays subtracting the single particle contribution,
\begin{equation}
    \boldsymbol{F}_j^{(2)} = \sum_{k  \neq j}\boldsymbol{f}_{jk}-\boldsymbol{f}_j \label{pair}
\end{equation}
where $\boldsymbol{{f}}_j$ is the 1-interaction force on a single particle. If the beam is a plane wave, then $\boldsymbol{f}_j=0$ in the transverse plane since $\partial_{\boldsymbol{q}}\mathbf E_0(\mathbf r_j)=0$. Therefore, as shown in the Supplementary Information, Eq. \Ref{pair} in a plane wave would only involve the 2-particle interaction. The 2-interaction force includes the optical binding force\cite{dholakia_colloquium_2010} as well as possible non-reciprocal forces on the center of mass on each dimer\cite{yifat_reactive_2018,peterson_controlling_2019,sukhov_actio_2015}.

The force described in Eq. \Ref{pair} only accounts for the 2-interaction contribution to the force on particle j due to particles k and does not account for multi-particle interactions (3-interaction and higher terms); i.e., the so called N-body contributions associated with $\alpha^3$ and higher order terms\cite{parker_symmetry_2025,chaumet_physical_2000,lakhtakia_macroscopic_1990,peterson_electrodynamic_2024}. We previously determined\cite{peterson_electrodynamic_2024} that the contribution of multi-particle scattering to the total electric field at single nanoparticles in small (up to 6 nanoparticle) OM arrays could be enhanced by as much as $40 \%$ over the incident field, $\boldsymbol{E}_0$. Therefore, the field and intensity gradients in 3 or more particle systems could be significantly different from the result of Eq. \Ref{pair}; that only considers the second-order 2-interaction term in the expansion of matter-radiation interactions. 

We can obtain insight into the N-body/N-interaction contributions to the force on the individual nanoparticles by calculating the 1-interaction, $\boldsymbol{F}_\text{{self}}(\boldsymbol{r}_j)$, and 2-interaction, $\boldsymbol{F}_{\text{(2),j}}$, forces on particle j obtained by Eqs. \Ref{alpha2},\Ref{pair} and comparing them to $\boldsymbol{F}_{\text{total,j}}$ on particle j in the presence of all neighboring nanoparticles. The purpose of this calculation is to explicitly obtain the 1-interaction and 2-interaction forces acting on each individual particle. The difference (discrepancy) in the sum of their force vectors and the total force  $\boldsymbol{F}_{\text{total,j}}$ are the N-interaction force(s), $\boldsymbol{F}_{\text{j}}^{(N)}$ on particle j. These force vectors are shown in Figure 5.

As described in the Supplementary Information, we performed EDLD simulations of the 8 Ag nanoparticle OM system to sample different configurations of the kite isomer and then grouped these into 4 averaged structures: (i) At the transition state ($d_1 = d_2$) for pseudorotation (with $D_4$ point group symmetry); (ii) At the peak of the probability distribution (termed $d_1 < d_2$) defined in Figure S11 in the Supplementary Information; (iii) less extended (termed $d_1 \leq d_2$) than the peak probability; (iv) and more extended (termed $d_1 \ll d_2$). Figure 5a-d shows these structures ordered from transition state (Figure 5a) to most extended (Figure 5d) with force vectors calculated in a Gaussian electromagnetic field (beam). The transition state (Figure 5a,e) is an unstable (saddle point) configuration while the "peak" (Figure 5c,g) structure is a local metastable configuration. The total force vectors $\boldsymbol{F}_{\text{total,j}}$ for the transition state structure and the "peak" structure (Figure 5a,c) are purely azimuthal. The total forces on the other structures (Figure 5b,d) are more complex and include radial components.

Figures 5e-h show that the 1-interaction forces (blue arrows) - the intensity gradient force on each nanoparticle - point radially inward towards the Gaussian beam's center, which is located at the origin. The 2-interaction forces (red arrows) partial offset the 1-interaction intensity gradient forces. However, summing the blue and red vectors would not equal the total force vectors (black) on each nanoparticle. The difference(es) between vector sum of the 1- and 2-interaction forces and the total force on each nanoparticle represents the N-interaction forces on that nanoparticle arising from multiple scattering from other particles. The green vectors in Figures 5e-h show that these N-interaction forces are the source of non-conservative orientational (azimuthal) motion of the OM array\cite{parker_optical_2020, han_crossover_2018,han_phase_2020,nan_creating_2022}. (Note that the N-interaction forces (green vectors) are nearly the same as the total forces (black) on the central 4 nanoparticles and are difficult to distinguish.) See the Supplementary Information and Figure S11 for analogous results obtained in a plane wave incident field where the 1-interaction gradient forces are absent.

\begin{figure}[H]	
	\centering
	\includegraphics[width=1\textwidth]{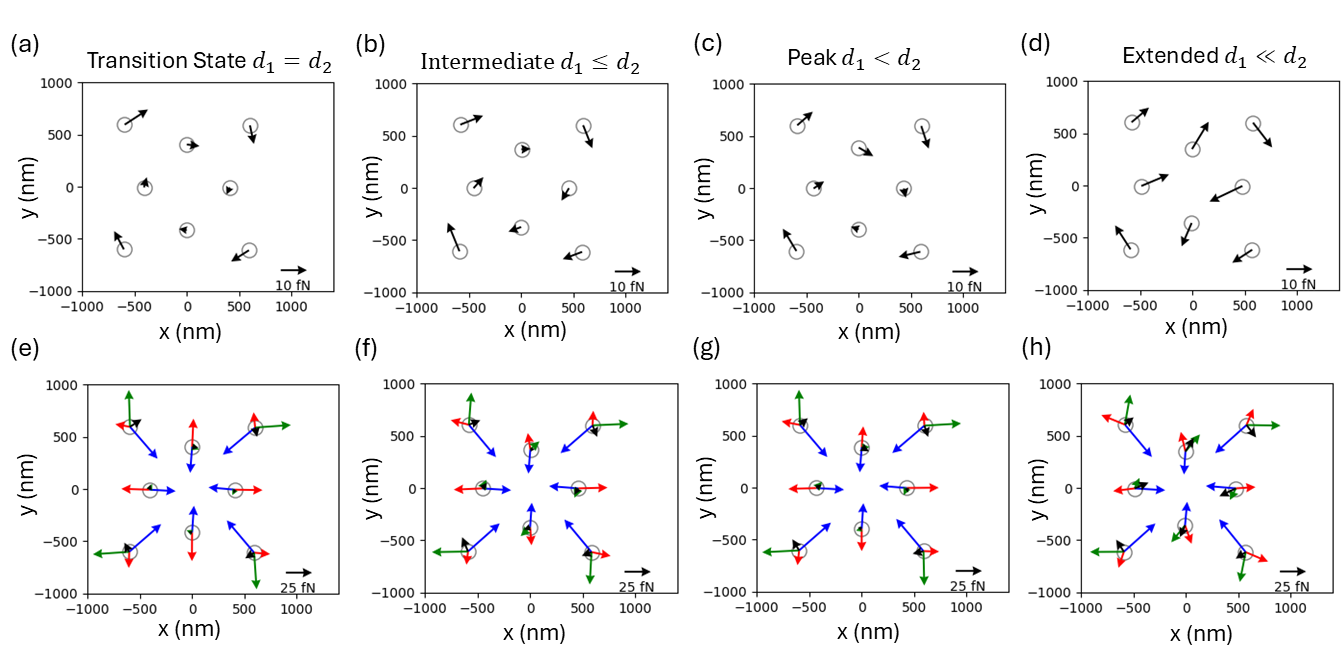}
	\caption{Total forces and force decomposition for various configurations for the kite isomer. (a)-(d) Force calculations of total force (black arrows) on each particle for 4 different kite structures. Each structure is an average of an ensemble of kite structures from a single EDLD simulation trajectory selected with different ranges of $d_1 - d_2$ values: (a) $d_1 = d_2 \pm 10 nm$; (b) $d_1 \leq d_2$ (90 nm to 110 nm);  (c) peak $d_1 < d_2$ (190 nm to 210 nm); (d) $d_1 \ll d_2$ (290 nm to 310 nm). (e)-(h) Vectors of total force and decomposed components on each nanoparticle of the structures in (a)-(d); total force (black arrows), intensity gradient force (blue arrows), the 2-interaction force (red arrows) and the N-interaction force (green arrows). Note that the scale bars are 10 fN in (a) - (d) and 25 fN in (e) - (h). See Supplementary Information for more details on the force calculations.} 
	\label{fig:5}
\end{figure}

 Our focus is to obtain qualitative insights from these results rather than dwelling on precise magnitudes of all the various 1-, 2- and N-interaction contributions. It is clear from the results in Figure 5 that forces on the nanoparticles associated with higher-order interactions, which we term N-interaction forces (green vectors), are significant in the various structures of the kite isomer. The N-interaction forces for the transition state structures in a Gaussian beam (Figures 5a,d) are the most simple to interpret; the N-interaction contribution to the force is the main origin of the azimuthal force. This is consistent with our previous work on negative torque\cite{han_crossover_2018} and creating optical matter machines\cite{parker_optical_2020} in highly symmetric (e.g., $D_6$) OM structures. The previous work was explained in terms of collective scattering resonances (i.e., lattice resonances) decomposed into spherical harmonic components of the far-field scattered EM field. High symmetry OM structures can manifest large angular momentum (e.g., 7 units of angular momentum for a 6-fold symmetric structure) in their scattered fields creating large non-conservative azimuthal forces on the nanoparticle constituents.\cite{parker_optical_2020}. In the present case, the forces at the $D_4$ symmetric transition state for pseudorotation of the OM kite system are due to analogous collective far-field scattering creating an azimuthal force that results in rigid body rotation. Supplementary Figure S8 shows that the kite isomers rotate; Figure S8 plots the raw (not rotation-reset) position data for both the experimental and simulation trajectories. The "peak" structure (Figure 5c,g) and the other two structures (Figure 5b,f and 5d,h) have $D_2$ symmetry so their collective scattering also causes a non-reciprocal azimuthal N-interaction force in addition to other non-azimuthal forces.

The other main conclusion to draw is that the N-interaction forces offset the inward-directed 1-interaction (intensity gradient) forces acting on the outer 4 nanoparticles to maintain the "expanded" kite structure. Therefore, \textit{the N-interaction forces are essential for the kite isomer to even exist as a structure that is not based on a trigonal lattice!}

\subsection{3.6 Particle motions for inter-state transitions of the 8-particle
OM system.}
 Optical matter systems formed with circularly polarized light typically arrange in trigonal lattice patterns that maximize the number of nanoparticle constituents located one optical binding distance apart. OM structures formed in linearly polarized focused laser beams generally deviate from cylindrical symmetry and have asymmetric shapes.\cite{nan_dissipative_2018} Figure 1 shows that the 8-NP OM system exists as multiple isomers, with the kite isomer deviating from a trigonal lattice. Figure 3f characterizes the first passage time distributions of the kite state transitioning to other states showing that transitions between the states (isomers) where particles are located on a trigonal lattice\cite{chen_raman_2024} occur more rapidly than transitions between these isomers and the kite state.  The relative probability of transitioning to any given configuration is related to the number of correlated or simultaneous particle displacements required for the transition. The most common transitions involving only 1 or 2 particle displacements or short sequences of single particle displacements. 
 
 Figure 6 shows sequences of dark-field microscopy images of 8-NP OM isomers and particle motions required for transitions. An isomerization that occurs frequently in the 8-particle OM system is between the "sphinx" state and the "spaceport" state. An example of this is shown in the first row of Figure 6. The NP constituents in both of these states are located on a trigonal lattice. The transition consists of few particles relocating to new sites on a trigonal lattice. These transitions are common since a small positional fluctuation of one particle from one site toward another can induce the displacement of its neighbors, allowing the isomerization to occur as a sequence of events without requiring simultaneous multiple particle events. 
 
 Transitions to and from the kite state are more complicated, perhaps because particles in the kite state are not located on a trigonal lattice. Most transitions from the other states (isomers) to kite and vice versa involve (concerted) many particle motions. Analysis of these transitions shows that they can occur over multiple frames of 450 fps experimental dark-field microscopy videos with unstable intermediate configuration(s) that are not on a trigonal lattice and are not the kite structure (Figure 6 rows b,c). Some transitions to and from the kite isomer occur more quickly; the transition can occur faster than the experimental time resolution of about 2 ms (450 fps). Furthermore, not every transition between a trigonal lattice isomer and kite has the same mechanism. For example, there exist multiple transition pathways between kite and "teardrop" (Figure 6b and 6c), and so there appear to be multiple transition states the system passes through between the two isomers. The most common transition to the kite state is from the "teardrop" state, see Figure 6 rows b,c. We have observed this transition occurring through multiple different pathways (SI Figure). The first involves many small displacements, often of all 8 constituents. Another pathway involves large amplitude NP displacements of a few (3-4) nanoparticles. 
 
 Both of these pathways involve the concerted motion of many nanoparticles which we believe explains why the transition into and out of kite is less frequent. Large fluctuations are less common than small ones, and the coordinated fluctuations of many NPs (6-8) is less common than the coordinated fluctuation of a few NPs (2-3) that we observe in the case of isomers with trigonal lattice structures. The reaction rates of the latter type of  transitions are within an order of magnitude of the time scale of data collection (450 fps). 

\begin{figure}[H]	
	\centering
	\includegraphics[width=1\textwidth]{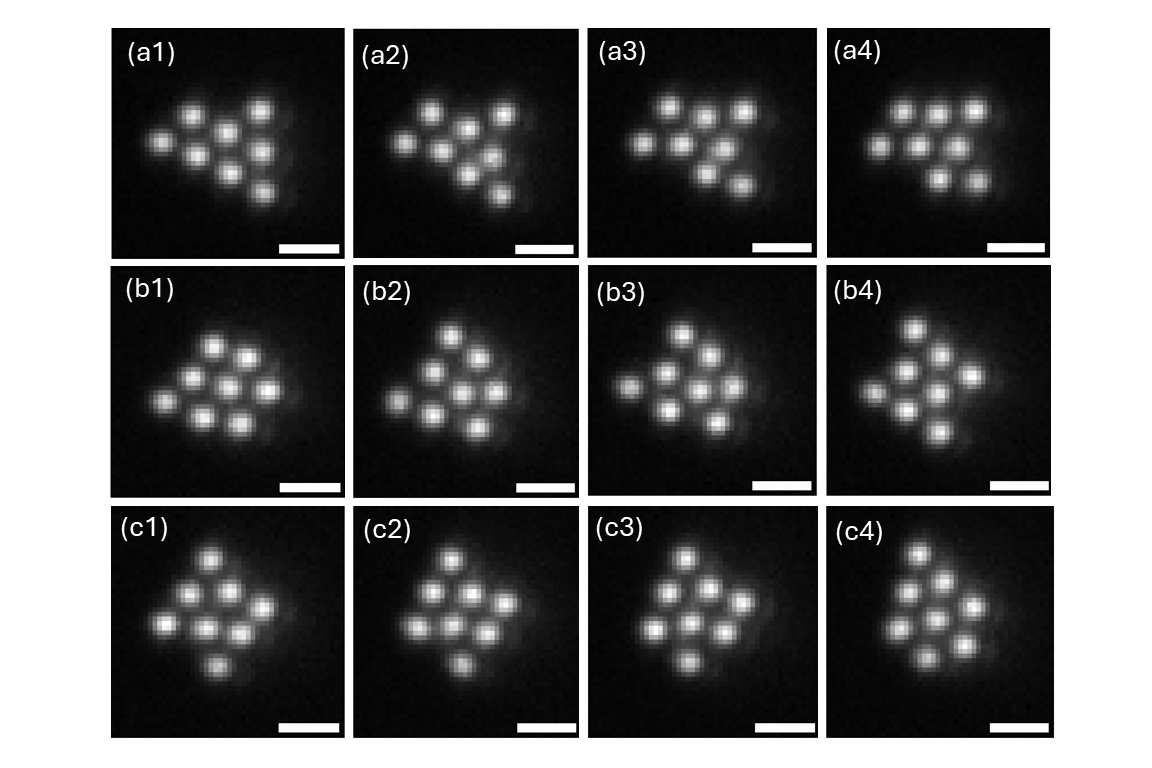}
	\caption{Three sets of consecutive frames of dark field microscopy videos of the 8-particle OM system (450 fps) documenting the single or collective particle motions associated with different isomerization pathways.  (a1-a4) Transition from sphinx to spaceport. (b1-b4) Transition from teardrop to kite. (c1-c4) Transition from kite to teardrop. Scale bars are 1 $\mu m$ }
	\label{fig:6}
\end{figure}

\section{Discussion and Conclusions}
Optical matter systems are analogous to chemically bonded systems. While OM systems do not manifest quantum mechanical chemical bonding, they exhibit similar phenomenon. For example, collective motions of OM systems, described as principal component (PC) modes, are analogous to vibrational normal modes, but the PC modes have been developed for overdamped non-conservative systems.\cite{chen_data-driven_2021}. The PC modes can be interpreted as a set of orthogonal collective modes that characterize the independent degrees of freedom of the OM system.\cite{chen_pseudorotation_2025} Principal component analysis (PCA) diagonalizes the covariance matrix of the OM particle coordinates to define a linear transformation into a basis of non-local collective modes. \cite{chen_data-driven_2021} Our recent paper\cite{chen_pseudorotation_2025}, and the present paper, extended these ideas to OM systems that do not have trigonal lattice-based structures. 

This paper characterizes the inter- and intra-state (isomer) transitions of an 8 particle optical matter system formed with 150 nm diameter Ag nanoparticle constituents in both experiment and simulation. We have shown the existence of a surprisingly stable "kite" isomer where nanoparticle constituents are not located on a trigonal lattice. This isomer manifests pseudorotation in two dimensions, a phenomenon that had previously only been studied in three dimensional molecular systems.\cite{berry_correlation_1960,cass_illustration_2005}  We showed that PCA Mode 3 that we reported in a related simulation paper\cite{chen_pseudorotation_2025} is a suitable reaction coordinate for the pseudorotation. This claim is supported by the committor analysis and the anti-correlation of the quadrilateral aspect ratio and $d_1$ - $d_2$ of kite isomer structures.  Changing the Debye screening length affects both inter- and intra-state dynamics of the 8-particle OM system. Both simulations and and experiments reveal that pseudorotation occurs more frequently when the Debye screening length is shorter (higher ionic strength) and that changing the Debye screening length changes the probability of the OM system being in the kite isomer (as well as all other isomers). 

\subsection{4.1 Stability and Dynamics of the isomers of the 8-particle
OM system.}

The transitions between trigonal lattice-based isomers are frequent and can occur rapidly (i.e., on the order of 1-10 ms), as small magnitude positional fluctuations of individual nanoparticles lead to rearrangements of the nanoparticle constituents onto different trigonal lattice sites. To first approximation, the stability of an isomer is related to the number of near-neighbor constituents located at a separation of one optical binding distance; it is tempting to suggest that these configurations represent local minima in the "potential energy surface" of the 8-NP OM system.\cite{yan_potential_2014}. The problem with this picture is that unlike molecular systems we do not have a fixed potential energy function for OM systems since they have both conservative and non-conservative forces and the system's energy depends on the positions of all the particles and cannot be factorized into sums of pair interactions\cite{chen_power_2024}. For these reasons, OM systems have to be described in terms of forces not potential energy functions and that N-body (N-interaction) forces, as presented in Section 3.5 and Figure 5, are important.

Perhaps surprisingly, the kite state (isomer) is significantly more stable than all other states once the system fluctuates into it. As shown in Figure 3f, the first passage time of the kite isomer transitioning to another isomer (state) is significantly longer than the inter-isomer transitions of its trigonal lattice-based counterparts. This longer first passage time indicates a larger activation energy into and out of the kite isomer. 

{From the ratio of the decay constants of the first passage time probability from Figure 1g, we estimate the activation energy of transitioning away from the kite state to be 2 times larger than the activation energy associated with transitions away from teardrop and potentially up to 4 times larger than transitions away from other trigonal lattice isomers. The large activation energy into and out of the kite isomer results from the collective many particle rearrangements required for the transitions.}

\subsection{4.2 Importance of N-body, N-interaction forces.}

The kite isomer of an OM system is a curious structure and the reason for its stability relative to that of the other 8-NP OM isomers is not immediately obvious. Previous studies of optical matter systems of different particle numbers ranging from 3-40 nanoparticle constituents in a circularly polarized beam find that almost all stable structures form with constituents located on a trigonal lattice.\cite{parker_optical_2020,chen_data-driven_2021,han_phase_2020} This behavior can be explained by extending the concept of an ideal optical binding distance \cite{yan_potential_2014} of a two particle optical matter system to larger many particle OM systems; i.e., the minimum energy configuration is the one that maximizes the number of nearest neighbor separations to be one optical binding distance apart.\cite{peterson_electrodynamic_2024} In our case that is $\sim$600 nm for an 800 nm trapping wavelength with water as the medium. There is also a local energy minimum at 2 optical binding distances ($\sim$1200 nm). The pairwise distribution functions found in Figure 1e-h show that isomers with particles on a trigonal lattice maximize the number of particles with these spacings, with teardrop having 14, while the kite isomer has fewer interparticle separations at these "ideal" distances. This fact suggests that the kite isomer would be higher energy than those isomers that form trigonal lattice structures, and are therefore less likely to occur. In non-equilibrium steady state conditions, the relative probabilities of the OM system being in a particular isomer are related to their relative interaction energies. Indeed, Figure 1 shows that the kite isomer is significantly less probable ($\sim 10\%$ total probability) than isomers with NPs on a trigonal lattice ($\sim 65\%$).

Since the kite structure deviates from a trigonal lattice, its significant relative stability, deduced from a first passage time analysis, is surprising. So why is the kite state stable if it doesn't adhere to this principle? The focused beam width and the strength of the inwards phase gradient of the trapping beam are factors that affect the stability of the kite state. In our first measurements of the 8-particle OM system created in a tightly focused beam it was found that the kite state was significantly less probable than the trigonal lattice states (0.19\% in 18 M$\Omega$ $H_2O$ with a tightly focused beam and 4\% in 18 M$\Omega$ $ H_2O$ with a less focused beam). So, the experimental conditions certainly affect the stability of the kite isomer relative to the other 8-NP isomers. 

However, the conclusion to draw from Figure 5e-h and Figure S11 is that the stability of the kite isomer for a particular set of experimental conditions (specifically, a relatively loosely focused Gaussian beam and 0.6 mM ionic strength) arises from the N-interaction forces that offset the inward-directed 1-interaction (intensity gradient) forces, particularly on the outer 4 nanoparticles since the 2-interaction (pairwise optical binding) forces (Figure 5) don't stabilize the kite isomer by themselves. These N-interaction forces maintain the "expanded" kite structure and are essential for the kite isomer to even exist as a structure that is not based on a trigonal lattice! Previous work has shown that symmetry breaking in OM systems leads to N-body forces in OM systems\cite{parker_symmetry_2025}. Therefore, the non-equilateral particle arrangements in the kite structure and other broken symmetries in the 8-particle OM system contribute to the N-interaction forces. We are developing new simulation code to determine these force components on an term-by-term basis (c.f., Eq. \Ref{total}) and will present the results in a future manuscript. 

\subsection{4.3 Pseudorotation in 2-D vs. 3-D systems}

We end by returning to the motivating idea of the Introduction - that pseudorotation of the Ag nanoparticles in the OM system studied here occurs in 2-D whereas it involves 3-D motions of atoms in molecular systems. Since chemical bonding in polyatomic molecules almost always involves non-planar structures, the bonding involves (molecular) orbitals that are 3-dimensional; e.g., trigonal bipyramidal hybridization and $D_{3h}$ symmetry or in square pyramidal molecules \cite{solomons_organic_2016,berry_correlation_1960,cass_illustration_2005,hargittai_r_2023,chen_cavity-enabled_2022}. Optical matter systems involve the exchange of photons (Bosonic particles) with interactions determined by the scattering of EM fields from the nanoparticles.\cite{peterson_electrodynamic_2024} We previously determined that accurate EDLD simulations of OM systems with metallic nanoparticle constituents using Generalized Multi-particle Mie theory (GMMT) only requires including dipole and quadrupole terms in the far-field scattering for converged energetics of interaction. \cite{parker_collective_2020}. Since the dipolar term is dominant and since that scattering is isotropic about the induced dipolar axis, the strongest inter-particle interactions will occur for parallel dipoles in a plane \cite{dholakia_colloquium_2010}. Our experimental setup (see Supplementary Figure S1) involves a balance of axial forces on the nanoparticles; i.e., a balance between incident radiation pressure and Coulomb repulsion with the nearby electrostatically charged glass coverslip of the fluid sample cell. Therefore, the only stable plane for the nanoparticles is a (horizontal) plane perpendicular to the propagation direction of the incident optical trapping laser beam. Also, unlike quantum mechanical orbitals that strongly constrain atomic motions in molecules, the electrodynamic interactions between the nanoparticles in OM systems are governed by interference of the incident and scattered fields. This is likely to be a lesser constraint on the possible motions than bonding in molecules thus allowing 2-dimensional pseudorotation in OM systems. That said, it would be of interest to create 3-D OM systems to determine if OM systems can also manifest 3-D pseudorotation.

\section{Acknowledgments}
We thank Dr. Curtis Peterson for preliminary experimental studies and suggesting the conditional probability representation shown in Figure 3. We thank the University of Chicago Department of Chemistry for the Helen Sellei-Beretvas/Georgine A. Moerke Fellowship. We acknowledge the DOD and ONR for a Vannevar Bush award that supported the early stage of this research.

\newpage
\bibliography{main}
\end{document}